\documentclass[12pt]{article}
\font\fff=eufm10 scaled \magstep1
\def\g{\hbox{\fff g}}
\begingroup 
\newtheorem{thm}{Theorem}[section]           
\newtheorem{lemma}[thm]{Lemma}
\newtheorem{proposition}[thm]{Proposition}
\newtheorem{theorem}[thm]{Theorem}
\newtheorem{corollary}[thm]{Corollary}
\newtheorem{definition}[thm]{Definition}
\newtheorem{notation}{Notation}

\newtheorem{remark}[thm]{Remark}             
\endgroup
\def\qed{\qquad\framebox[7pt]\medskip\noindent}

\newcommand\ad{{\rm ad}\,}
\def\half{{\textstyle\frac{1}{2}}}
\newenvironment{proof}{{\sl Proof:}\quad}{\hfill{\qed}\\ \noindent}
\renewenvironment{abstract}{\begin{quote}{\bf Abstract.\
}\small}{\end{quote}\bigskip}
\title{Representations of the Schr\"odinger algebra and Appell systems}

\author{Philip Feinsilver\thanks{\leftskip -1cm Department of Mathematics
Southern Illinois University 
 Carbondale, IL. 62901, U.S.A.}, 
Jerzy Kocik\thanks{\leftskip -1cm Department of Mathematics
Southern Illinois University 
 Carbondale, IL. 62901, U.S.A.}, 
and Ren{\'e} Schott\thanks{\leftskip -1cm IECN and LORIA
 Universit\'e Henri Poincar\'e-Nancy 1, BP 239, 
54506 Vandoeuvre-l\`es-Nancy, \hbox to 18pt{}France.}} 

\date{}

\begin{document}

\maketitle

\begin{abstract}
We investigate the structure of the Schr\"odinger algebra and
its representations in a Fock space realized 
in terms of canonical Appell systems. Generalized coherent states are
used in the construction of a Hilbert space of functions on which
certain commuting elements act as  self-adjoint operators. 
This yields a probabilistic interpretation of these operators as
random variables. An interesting feature is how the structure of the
Lie algebra is reflected in the probability density function.
A Leibniz function and orthogonal basis for the Hilbert space is found. Then
Appell systems connected with certain evolution
equations, analogs of the classical heat equation, on this algebra are computed.



\end{abstract}

\vfill
\pagebreak
\tableofcontents
\pagebreak

\section{Introduction}

The Schr\"odinger Lie algebra plays an important role in
mathematical physics and its applications 
(see, e.g., \cite{BHP, BR, BX, Ha, Ni}). 
Using the technique of singular vectors, a classification of the
irreducible lowest weight representations of this algebra is given in
\cite{DDM}. A main feature of the present paper is
a classification of the representations of the Schr\"odinger 
algebra in an alternative way based on the semidirect product
structure of the algebra.\\

We begin in \S2 with some interpretations of the notion of `Appell systems'.
Section 3 contains basic details of our approach to representations of the
Schr\"odinger algebra. In particular, we show how it is built and
determine a standard form. Some group calculations are done using a
matrix realization of the algebra.  
In \S4 we construct canonical Appell systems and find a family of probability
distributions associated to the
Schr\"odinger algebra that reflects its Lie algebraic structure. 
In particular, we see that the results of \cite{DDM} on
polynomial representations  based on lowest weight modules fit into
our picture. The details of the associated
Hilbert space comprise \S5. This starts with computing the Leibniz
function. We show how to recover the Lie algebra from the Leibniz
function and obtain an orthogonal basis for the Hilbert space. 
In the final section, 
we show how to construct Appell systems which provide solutions to generalized heat
equations on the Schr\"odinger algebra,
corresponding to classical two-dimensional real L\'evy processes.
\\

\section{Appell systems: some interpretations}
\label{sec:apsys}

There are three interpretations of the notion of `Appell systems':

\begin{enumerate}
\item Appell systems in the classical sense. These are considered
below from a more general viewpoint.

\item Canonical Appell systems associated to a Lie algebra. 
One uses the Lie algebra to construct a Hilbert space with the Appell
system as basis. See \S4.

\item General Appell systems on Lie groups. Here one uses the Lie
algebra and group structure as a `black box' into which a classical
stochastic process goes in and produces a `Lie response' --- typically
a process consisting of iterated stochastic integrals of the input
process (\cite{FS1, FS2}). In this paper, for simplicity, we restrict to
the abelian case. Appell systems provide solutions to evolution
equations related to the input stochastic process. See \S\ref{sec:apsyss}.
\end{enumerate}

Now we shall expand on the first point of view.\\

Considering the differentiation operator $D$, we may think of the space
of polynomials of degree not exceeding $n$ as the space of
solutions, ${\cal Z}_n$, to the equation $D^{n+1}\psi=0$. In this context 
an {\it Appell system} is defined to be a sequence of nonzero polynomials
$\{\psi_0,\psi_1,\ldots,\psi_n,\ldots\}$ satisfying:

\begin{enumerate}
\item $\psi_n\in{\cal Z}_n$, $\forall n\ge0$ 
\item $D\psi_n=\psi_{n-1}$, for $n\ge1$
\end{enumerate}

(Note that this differs slightly from the usual definition, cf.
\cite{FS1}, which has $D\psi_n=n\psi_{n-1}$.)
By analogy,  for any operator $V$, called the 
\emph{canonical lowering operator,}  we define a $V$-Appell system as follows.
Set $${\cal Z}_n=\{\,\psi\;\colon\; V^{n+1}\psi=0\,\}$$ for $n\ge0$. 
Then the {\em $V$-Appell space decomposition} is the system of 
embeddings ${\cal Z}_0\subset{\cal Z}_1\subset{\cal Z}_2\subset\ldots$, 
and a {\em $V$-Appell system} is a sequence of nonzero functions
$\{\psi_0,\psi_1,\ldots,\psi_n,\ldots\}$ satisfying:

\begin{enumerate}
\item $\psi_n\in{\cal Z}_n$, $\forall n\ge0$ 
\item $V\psi_n=\psi_{n-1}$, for $n\ge1$
\end{enumerate}

Typically, one starts with a `standard Appell system', such as
$\psi_n=x^n/n!$, for $V=D$. Then Appell systems are generated from the 
standard one via time-evolution. To accomplish this for $V$-Appell systems,
the symmetry algebra of $V$ comes into play.\\

If $V$ is an operator acting on a space of smooth functions, we define 
its {\em unrestricted symmetry algebra} to be the Lie algebra $\g(V)$ 
of vector fields, $X$, such that there exists an operator $\Lambda(X)$ in the
center of $\g(V)$ with 
$$ 
       [X,V]=\Lambda(X)\,V
$$  
If in every case 
we require $\Lambda$ to be multiplication by a scalar function, we shall talk of
the {\em restricted} symmetry algebra, as in \cite{DDM}.
If we consider only those $X$ for which $\Lambda(X)=0$, we have the {\em strict}
symmetry algebra $\g_0(V)\subset\g(V)$.  Clearly, $V\in\g_0(V)$.
Also, it is clear that $\g_0(V)$ contains the center of 
$\g(V)$.
\\

\begin{proposition}
The strict symmetry algebra contains the derived algebra of unrestricted
symmetries:
$\g'(V)\subset \g_0(V)$. That is, $Y\in \g'(V)$ implies $[Y,V]=0$.
\end{proposition}

\begin{proof}
  Let $[X_1,V]=\Lambda(X_1)\,V$ and $[X_2,V]=\Lambda(X_2)\,V$. 
  Then, by the Jacobi identity,
  \begin{eqnarray*}
    [[X_1,X_2],V] &=& [X_1, \Lambda(X_2)\,V]+[\Lambda(X_1)\,V, X_2] \\
                  &=& \Lambda(X_2)\,[X_1,V]+\Lambda(X_1)[V,X_2] \\
                  &=& \Lambda(X_2)\Lambda(X_1)\,V 
                      - \Lambda(X_1)\Lambda(X_2)\,V = 0
  \end{eqnarray*}
using the property that the $\Lambda$ operators are central.
\end{proof}

The relevance for $V$-Appell systems is this.

\begin{proposition}
The unrestricted symmetry algebra $\g(V)$ of an operator $V$
preserves the  Appell space decomposition 
${\cal Z}_0\subset{\cal Z}_1\subset{\cal Z}_2\subset\ldots$, 
that is, $X{\cal Z}_n\subset {\cal Z}_n$ for every $X\in\g(V)$. 
\end{proposition}

\begin{proof}
Write $[X,V]=\Lambda(X)\,V$ in the form $VX=(X-\Lambda(X))V$. Fix
$n\ge0$ and let $\psi\in {\cal Z}_n$. Since
$\Lambda(X)$ commutes with $V$, we have 
$V^{n+1}X\psi=(X-(n+1)\Lambda(X))V^{n+1}\psi=0$.
\end{proof}

New Appell systems are generated from a given one by the adjoint action 
of a group element generated by a `Hamiltonian' --- a
function of elements of the symmetry algebra. The structure of
the spaces ${\cal Z}_n$ is preserved, while the Appell systems provide
`polynomial solutions' to the evolution equation corresponding to the
Hamiltonian. Indeed, if $H$ is a function of operators in $\g(V)$, with
$H\psi_0=0$, then $h_n=\exp(tH)\psi_n$ will be an Appell system.
For each $n$, the function $h_n$ satisfies $u_t=Hu$, with $u(0)=\psi_n$. 
In the simplest situation where $H$ is a function of $D$ and the initial
Appell 
sequence is $\psi_n=x^n/n!$, different choices of $H$ yield many of the
classically important sequences of polynomials (with perhaps minor variations). 
\\

In the paper \cite{DDM}, a hierarchy of solutions to 
${\bf S}^{(p/2)}\psi=0$ is developed for the Schr\"odinger operator
$\bf S$. 
The representations discussed there can be viewed as $\bf S$-Appell
systems in the above sense. These correspond to
finite-dimensional representations of sl(2) in the standard form of
the Schr\"odinger algebra given below.
\\

\section{Schr\"odinger algebra}
\label{sec:envalg}

Referring to \cite{DDM} for details,
recall that the ($n=1$, centrally-extended) Schr\"odinger algebra ${\cal S}_1$
is spanned by the following elements (commented by their physical origins):
 
\begin{eqnarray*}
M   &\qquad& \hbox{\rm mass} \\
K   &\qquad& \hbox{\rm special conformal transformation} \\
G   &\qquad& \hbox{\rm Galilei boost} \\
D   &\qquad& \hbox{\rm dilation (not differentiation!)} \\
P_x &\qquad& \hbox{\rm spatial translation} \\
P_t &\qquad& \hbox{\rm time translation} \\
\end{eqnarray*}

which satisfy the following commutation relations, 
given here in the form of a matrix with rows and columns 
labelled by the corresponding operators
$$
\bordermatrix{
     & M & K    & G   & D     & P_x &    P_t \cr
M    & 0 & 0    & 0   & 0     &   0 &      0 \cr 
K    & 0 & 0    & 0   & -2\,K &  -G &     -D \cr 
G    & 0 & 0    & 0   & -G    &  -M &   -P_x \cr 
D    & 0 & 2\,K & G   & 0     &-P_x &-2\,P_t \cr 
P_x  & 0 & G    & M   & P_x   &   0 &      0 \cr 
P_t  & 0 & D    & P_x & 2\,P_t&   0 &      0 \cr 
}
$$

Note that elements $\{M, G, P_x\}$ span a Heisenberg-Weyl subalgebra, 
while $\{K, D, P_t\}$ span an sl(2) subalgebra. 
This fact, that the Schr\"odinger algebra is a semidirect product
$$
{\cal S}_1 \cong {\cal H} \oplus_s sl(2)
$$
is the basis for analyzing the representations of the Schr\"odinger
algebra. We continue with the $n=1$ case and indicate how the general
case $n>1$ goes at the end of the discussion of the standard form, since
the rotation generators, $J_{ij}$, do not appear in the case $n=1$.
\\

\subsection{Structural decomposition for Fock calculus}

In general, in order to construct representations, 
we first seek a {\em generalized Cartan decomposition} of the Schr\"odinger algebra
into a triple $\g={\cal P}\oplus{\cal K}\oplus{\cal L}$  
where $\cal P$ and $\cal L$ are abelian subalgebras, and 
$\cal K$ is a subalgebra normalizing both $\cal P$ and $\cal L$.
The main idea is that elements of ${\cal P}$ and $\cal L$ 
act as raising and lowering operators, respectively.
The possibility of finding a scalar product in which each element of  
$\cal P$ has a corresponding adjoint in $\cal L$ is important, since
we wish to construct a family of selfadjoint  operators that provide   
a family of commuting quantum observables or classical random variables in the 
probabilistic interpretation.  In many cases, this family arises 
by conjugating elements of ${\cal P}$ by a group element 
with a generator from ${\cal L}$.
This technique may be viewed as an extension of the Cayley transform 
for symmetric spaces. Notice that for this to work, the
subalgebras  $\cal P$ and $\cal L$ must be in one-to-one
correspondence --- the \emph{Cartan involution} in the theory of
symmetric spaces.
\\

The Schr\"odinger algebra $\g={\cal S}_1$ admits the following 
generalized Cartan decomposition:
\begin{equation}\label{RNL}
\{m,K,G\} \oplus \{D,P_x\} \oplus \{P_t\}
\end{equation}
Note however that  ${\cal P}$ and ${\cal L}$ cannot be put into 1-1
correspondence and therefore this is of no direct use for us.
\\

We will use instead the following decomposition
( cf. \cite[p. 31]{He}):

\begin{equation}
\label{eq:alaCartan}
           \underbrace{ \{K,   G   \}  }_{\cal P} 
    \oplus \underbrace{ \{M,   D   \}  }_{\cal K}    
    \oplus \underbrace{ \{P_t, P_x \}  }_{\cal L}
\end{equation}

$M$ acts here as a scalar $m$. We take $R_1=K$ and $R_2=G$ as raising
operators. Even though $P_x$ is not in ``Cartan's $\cal{L}$'', as in
equation (\ref{RNL}), we use it as 
the lowering operator dual to $G$, so take $L_1=P_t$ and $L_2=P_x$.
\\

Even though the decomposition (\ref{eq:alaCartan}) is not technically
a Cartan decomposition, it will lead to interesting results
for representations of the Schr\"odinger algebra.
\\

\subsection{A matrix representation and group calculations}

A 4-dimensional representation (see \cite{BPPW}) of the Schr\"odinger
algebra ($n=1$) is given by embedding into su(4). 
Let $X$ denote a typical element of the Lie algebra. Set, 
\begin{equation}
\label{eq:typical} 
X = a_1m+a_2K+a_3G+a_4D+a_5P_x+a_6P_t = 
\pmatrix{0& a_5& a_3& 2a_1 \cr 
         0& a_4& a_2& a_3 \cr 
         0& -a_6& -a_4& -a_5 \cr 
         0& 0& 0& 0 \cr }
\end{equation}

We will denote a typical group element according to the basis we have chosen 
by 
\begin{eqnarray*}
     &&g(A_1,A_2,A_3,A_4,A_5,A_6)=      \\
     &&\qquad\exp(A_1m)\exp(A_2K)\exp(A_3G)\exp(A_4D)\exp(A_5P_x)\exp(A_6P_t)
\end{eqnarray*}

The variables $\{A_i\}$ are \emph{coordinates of the second kind.}\
The group element corresponding to (\ref{eq:typical}) is
$$
g(A_1,A_2,A_3,A_4,A_5,A_6)=e^{-A_4}\cdot
\pmatrix{e^{A_4} & A_5e^{A_4}-A_3A_6 & A_3 & 2A_1e^{A_4}-A_3A_5 \cr
0 &e^{2A_4}-A_2A_6 &A_2 & -A_3e^{A_4}-A_2A_5\cr
0 &-A_6 & 1& -A_5 \cr
0 & 0&0& e^{A_4} \cr}
$$

From this we have

\begin{proposition}\label{prop:coords}
  Given in matrix form a group element $g$, we can recover the second-kind 
coordinates $(A_1,\ldots,A_6)$ according to

\begin{eqnarray*}
&A_1=-\frac12\,{\textstyle\left|\matrix{g_{13}&g_{14}\cr
      g_{33}&g_{34}\cr}\right|\over\textstyle g_{33}},  \quad
&A_2={g_{23}\over g_{33}},\quad
 A_3={g_{13}\over g_{33}},\\
&A_4=-\log(g_{33}),\quad
&A_5=-{g_{34}\over g_{33}},\quad
 A_6=-{g_{32}\over g_{33}}
\end{eqnarray*}
\end{proposition}

Referring to decomposition (\ref{eq:alaCartan}), we specialize
variables, writing $V_1,V_2,B_1,B_2$ for $A_2,A_3,A_6,A_5$ respectively.
Basic for our analysis is the partial group law:
$$ 
       e^{B_1 P_t+B_2 P_x}\,e^{V_1K+V_2G} = \ ?
$$ 
We will get the required results using the matrix representation 
noted above. The general elements of $\cal P$ and $\cal L$ are:

\begin{eqnarray*}
B_1 P_t+B_2 P_x &=&
\left (\begin {array}{cccc} 0&{\it B_2}&0&0\\\noalign{\medskip}0&0&0&0
\\\noalign{\medskip}0&-{\it B_1}&0&-{\it B_2}\\\noalign{\medskip}0&0&0&0
\end {array}\right ) 
\\
V_1K+V_2G &=&
\left (\begin {array}{cccc} 0&0&{\it V_2}&0\\\noalign{\medskip}0&0&{
\it V_1}&{\it V_2}\\\noalign{\medskip}0&0&0&0\\\noalign{\medskip}0&0&0&0
\end {array}\right )
\end{eqnarray*}

As the square of each of these matrices is zero, the exponential of each 
reduces to simply adding the identity. We find the matrix product
$$
e^{B_1 P_t+B_2 P_x}\,e^{V_1K+V_2G} 
  =\left( \begin {array}{cccc} 
   1&{\it B_2}&{\it V_2}+{\it B_2}\,{\it V_1}&{\it B_2}\,{\it V_2}\\
   \noalign{\medskip}
   0&1&{\it V_1}&{\it V_2}\\
   \noalign{\medskip}
   0&-{\it B_1}&1-{\it B_1}\,{\it V_1}&-{\it B_1}\,{\it V_2}-{\it B_2}\\
   \noalign{\medskip}0&0&0&1
   \end {array}\right)
$$

Applying Proposition \ref{prop:coords} to the matrix found above yields

\begin{proposition}
\label{prop:Leibnizformula}
In coordinates of the second kind, we have the Leibniz formula, 
\begin{eqnarray*}
&&g(0,0,0,0,B_2,B_1)\; g(0,V_1,V_2,0,0,0)= \\
&&\qquad g\bigg(\frac12\,
         {\frac {{ B_1}\,{{ V_2}}^{2}+2\,{ B_2}\,{ V_2}+{{ B_2}}^{2}{ V_1}}
                       {1-{ B_1}\,{ V_1}}},\;
         {\frac {{ V_1}}{1-{ B_1}\,{ V_1}}},\;
         {\frac {{ V_2}+{ B_2}\,{ V_1}}{1-{ B_1}\,{V_1}}},  \\
&&\qquad\qquad\qquad
         -\log (1-{ B_1}\,{ V_1}),\;
         {\frac {{ B_1}\,{ V_2}+{ B_2}} {1-{ B_1}\,{ V_1}}},\;
         {\frac {{ B_1}}{1-{ B_1}\,{ V_1}}}\bigg)
\end{eqnarray*}
\end{proposition}

In general, a \emph{Leibniz formula} is the group law for commuting
the $L$ operators past the $R$'s, in analogy to the classical formula
of Leibniz for derivatives.
\\

\subsection{Standard form of the Schr\"odinger algebra}

Now we show the internal structure of the Schr\"odinger algebra ($n=1$).

\begin{remark}\rm \label{rem:1}
  Note that we work in enveloping algebras throughout, so our
  calculations are based on relations in an associative algebra. 
  In   particular, we often use
\begin{equation}\label{eq:commdv}
 [A,BC]=[A,B]C+B[A,C]\quad{\rm and}\quad
   [A,B^2]=[A,B]B+B[A,B]
\end{equation}
\end{remark}

\begin{definition}\rm
Denote the basis for a standard Heisenberg-Weyl (HW) algebra, 
${\cal H}=\hbox{\rm span}\,\{P,X,H\}$, satisfying
$$ 
            [P,X]=H,\quad [P,H]=[X,H]=0 
$$
A representation of HW-algebra such that $H$ acts as the scalar $m$ times
the identity 
operator will be denoted as $m$-HW algebra.
\end{definition}

\begin{definition}\rm
 Denote the basis for a standard sl(2) algebra, ${\cal K}$, by
$\{L,R,\rho\}$, satisfying
$$ 
         [L,R]=\rho,\quad [\rho,R]=2R,\quad [L,\rho]=2L
$$
We write ${\cal K}\colon=\{L,R,\rho\}$.
\end{definition}

The following Lemma is well-known.
It follows readily from the equations in remark \ref{rem:1}
(also see calculations below).

\begin{lemma}
Given an $m$-{\rm HW} algebra, setting 
$$ L=\frac{1}{2m}\,P^2,\quad \rho=\frac{1}{m}\,XP+\frac12,\quad
R=\frac{1}{2m}\,X^2$$
yields a standard {\rm sl(2)} algebra.
\end{lemma}

Now for our first main observation, which follows immediately from the 
commutation rules for the Schr\"odinger algebra.

\begin{thm} 
\label{thm:HW}
{\rm (HW form of the Schr\"odinger algebra)}
Given an $m$-{\rm HW} algebra, setting
$$ m=H,\,K=\frac{1}{2m}\,X^2,\,G=X,D=\frac{1}{m}\,XP+\frac12,\,
P_x=P,\,P_t=\frac{1}{2m}\,P^2$$
yields a representation of ${\cal S}$.  
\end{thm}

And the main theorem, which gives the standard form.

\begin{thm}
\label{thm:stdform}
{\rm (Standard form of the Schr\"odinger algebra)} 
Any representation of the Schr\"odinger algebra 
${\cal S}=\hbox{\rm span}\,\{m,K,G,D,P_x,P_t\}$
contains a standard {\rm sl(2)} algebra 
${\cal K}_0=\hbox{\rm span}\,\{L_0, R_0, \rho_0\}$ such that,
with the $m$-{\rm HW} algebra ${\cal H}=\hbox{\rm span}\,\{P_x,G,m\}$ from the
given representation of ${\cal S}$, 
the {\rm sl(2)} subalgebra is of the form
$$ 
K=R_0+\frac{1}{2m}\,G^2,\quad
D=\rho_0+\frac{1}{m}\,GP_x+\frac12,\quad
P_t=L_0+\frac{1}{2m}\,P_x^2
$$
where ${\cal K}_0$ commutes with ${\cal H}$.\\

Conversely, given any $m$-{\rm HW} representation, use it
for ${\cal H}:=\{P_x,G,m\}$. Now take any
{\rm sl(2)} algebra commuting with ${\cal H}$, 
and form the direct product with the standard
{\rm sl(2)} algebra constructed from ${\cal H}$ by the Lemma.
Then this yields a representation of ${\cal S}$.
\end{thm}

\begin{proof}
  The converse is clear by construction and our previous
  observations. What must be checked is that given a representation of 
  ${\cal S}$, setting
$$
R_0=K-\frac{1}{2m}\,G^2,\,
      \rho_0=D-\left(\frac{1}{m}\,GP_x+\frac12\right),\,
       L_0=P_t-\frac{1}{2m}\,P_x^2
$$
yields an sl(2) algebra that commutes with ${\cal H}$. From equation
(\ref{eq:commdv}), we have
$$[L_0,G]=[P_t,G]-\frac{1}{2m}\,[P_x^2,G]=P_x-P_x=0$$
and similar relations for $R_0$ and $\rho_0$ show that
${\cal K}_0$ commutes with ${\cal H}$. Now, using remark \ref{rem:1}, 
we note these relations
\begin{eqnarray*}
  [P_x^2,K] &=& P_xG+GP_x = 2GP_x+m \\
\lbrack GP_x,K] &=& 0\cdot P_x+G\cdot G = G^2 \\
\lbrack P_t,GP_x] &=& P_x^2
\end{eqnarray*}

Thus, using the fact that $[{\cal K}_0,{\cal H}]=0$, we have
\begin{eqnarray*}
  [L_0,R_0] &=& [L_0,K-\frac{1}{2m}\,G^2]\\
               &=& [P_t-\frac{1}{2m}\,P_x^2,K]+
                                    [L_0,-\frac{1}{2m}\,G^2]\\
               &=& D - \frac{1}{m}\,GPx-\frac12 = \rho_0
\end{eqnarray*}
while
\begin{eqnarray*}
[\rho_0,R_0] &=& [\rho_0,K-\frac{1}{2m}\,G^2]=[\rho_0,K]\\
             &=& [D,K] - \frac{1}{m}\,[GP_x,K]\\
             &=& 2K-\frac{1}{m}\,G^2=2R_0
\end{eqnarray*}
and
\begin{eqnarray*}
  [L_0,\rho_0] &=& [P_t-\frac{1}{2m}\,P_x^2,\rho_0] \\
                    &=& [P_t,D]-[P_t,\frac{1}{m}\,GP_x]\\
                    &=& 2P_t-\frac{1}{m}\,P_x^2 = 2L_0
\end{eqnarray*}
which completes the proof.
\end{proof}

\begin{remark}
The theorem, extended to include rotations,
holds also for $n>1$, where we use ${\cal K}_0$ 
spanned by
$$
  L_0  =\frac{1}{2m}\,\sum_i P_i^2,\quad
  R_0  =\frac{1}{2m}\,\sum_i G_i^2,\quad
\rho_0 = \frac{1}{m}\,\sum_i G_iP_i+\frac{n}{2} 
$$
and for the rotations,
$$ 
     J_{0,ij} = J_{ij} - \frac{1}{m}\,(G_iP_j-G_jP_i)
$$
with the $J_0$ rotations commuting with ${\cal H}$.\\
\end{remark}

As an application of Theorem \ref{thm:HW}, consider the special
realization, with scalar $M=m$ and $x$ denoting multiplication by the
variable $x$,

\begin{equation}
\label{eq:special}
  G  =mx,\quad P_x=\frac{d}{dx},\quad
  P_t=\frac{1}{2m}\,\frac{d^2}{dx^2},\quad
  K  =\frac{mx^2}{2},\quad D=x\,\frac{d}{dx}+\frac12
\end{equation}

In this realization, acting on the function identically equal to 1, we 
have $P_t1=P_x1=0$, and $D1=1/2$. 
Applying a group element to the function $1$, we find
$$
       g(A_1,A_2,A_3,A_4,A_5,A_6)\,1
              =\exp\left(A_1m+A_2\frac{mx^2}{2}+A_3mx+A_4/2\right)
$$
Clearly,  $f(x)1$ can be identified with the function $f(x)$ itself.
Now apply the Leibniz formula, Proposition \ref{prop:Leibnizformula}, to find

\begin{corollary}
\label{cor:leib}
The differential realization of the Schr\"odinger algebra ${\cal S}_1$
has the following ``partial group law"
\begin{eqnarray*}
&&\exp\left(\frac{B_1}{2m}\frac{d^2}{dx^2}+B_2\frac{d}{dx}\right)
            \exp\left(V_1\frac{mx^2}{2}+V_2\,mx\right)\\
&&\qquad
   =\exp\left(\frac{V_1}{1-B_1V_1}
             \frac{mx^2}{2}+\frac{V_1B_2+V_2}{1-B_1V_1}\,mx\right)\\
&&\qquad\qquad
   \times\,(1-B_1V_1)^{-1/2}
           \exp\left(\frac{m}{2}\,{\frac {{ B_1}\,{{ V_2}}^{2}
            +2\,{ B_2}\,{ V_2}+{{ B_2}}^{2}
             {V_1}}{1-{ B_1}\,{ V_1}}}\right)
\end{eqnarray*}
\end{corollary}

\section{Canonical Appell systems for the Schr\"odinger algebra}

Now to construct the representation space and basis --- the canonical
Appell system. To start, define a vacuum state $\Omega$ such that, for constants $m$
and $c$,
\begin{eqnarray*}
    K\Omega = K\Omega  &\qquad&  G\Omega = G\Omega  \\ 
    P_t\Omega=0\quad   &\qquad&  P_x\Omega=0       \phantom{\biggm|} \\ 
    M\Omega=  m\Omega  &\qquad&  D\Omega=c\Omega   
\end{eqnarray*}

\begin{notation}\rm
The standard form (cf. Theorem \ref{thm:stdform}) gives 
$D=\rho_0+(1/m)GP_x+1/2$, which shows that $\rho_0\Omega =(c-1/2)\Omega$.
Hence in the following we denote $c-1/2$ by $\dot c$. 
\end{notation}

The (commuting) elements $K$ and $G$ of $\cal P$ can be used to 
form basis elements 
$$
          |jk\rangle=K^jG^k\Omega\,, \qquad\quad j,k\geq 0
$$ 
of a Fock space ${\cal F}=\hbox{span}\,\{\;|jk\rangle \}$
on which $K$ and $G$ act as raising operators,
while $P_t$ and $P_x$ act as lowering operators. 
\\

\subsection{Adjoint operators and Appell systems}

The goal is to find an abelian subalgebra spanned by some 
selfadjoint operators acting on the representation space just constructed.
Such a two-dimensional subalgebra can be obtained by an appropriate ``turn''
of the plane $\cal P$ in the Lie algebra,
namely via the adjoint action of the group element formed by exponentiating $P_t$.  
The resulting plane, ${\cal P}_{\beta}$ say, is abelian and is spanned by
\begin{eqnarray}\label{eq:xs}
X_1 = e^{\beta P_t}Ke^{-\beta P_t} 
                      &=&\exp(\ad \beta P_t) K= K+\beta D+\beta^2 P_t\cr 
X_2 = e^{\beta P_t}Ge^{-\beta P_t}
                      &=& G+\beta P_x
\end{eqnarray}

Next we determine our canonical Appell systems.
We want to compute $\exp(z_1 X_1+z_2X_2)\Omega$.  
Setting $V_1=z_1$, $V_2=z_2$, $B_1=\beta$, and $B_2=0$
in Proposition \ref{prop:Leibnizformula} yields
\begin{eqnarray}
\label{eq:gaction}
&&e^{z_1X_1}e^{z_2X_2}\Omega=e^{\beta P_t}e^{z_1K}e^{z_2G}
      e^{-\beta P_t}\Omega = e^{\beta P_t}e^{z_1K}e^{z_2G}\Omega \nonumber \\
&&\quad
   =\exp({z_1K\over 1-\beta z_1})\exp({z_2G\over 1-\beta z_1}) 
        (1-\beta z_1)^{-c}\exp\Big({m\over 2}\,
        {\beta z_2^2\over 1-\beta z_1}\Big)\Omega
\end{eqnarray}

To get the generating function for the basis $|jk\rangle$, 
set in equation (\ref{eq:gaction}) 

\begin{equation} 
\label{eq:vz}
  v_1={z_1\over 1-\beta z_1}\,,\qquad 
  v_2={z_2\over 1-\beta z_1} 
\end{equation}

Substituting throughout, we have

\begin{proposition}
\label{prop:genfun}
The generating function for the canonical Appell system,
$\{|jk\rangle=K^j G^k\Omega\}$ is
\begin{eqnarray*}
   &&e^{v_1K}e^{v_2G}\Omega= \\ 
   &&\exp(x_1\,{v_1\over 1+\beta v_1})
     \exp(x_2\,{v_2\over 1+\beta v_1}) 
     (1+\beta v_1)^{-c}
     \exp\Big(-{m\beta\over 2}\,{v_2^2\over 1+\beta v_1}\Big)
\end{eqnarray*}

where we identify $X_1\Omega=x_1\cdot 1$ and $X_2\Omega=x_2\cdot 1$ in
the realization as functions of $x_1,x_2$.
\end{proposition}

With $v_2=0$, we recognize the generating function for the Laguerre
polynomials, while $v_1=0$ reduces to the generating function for
Hermite polynomials. 
This corresponds to the results of Section 4 of \cite{DDM}.
\\

From the exponentials $\exp(z_iX_i)$, equation (\ref{eq:gaction}), we
identify as operators
$z_1={\partial / \partial x_1}$ and
$z_2={\partial / \partial x_2}$.
Using script notation for the $v_i$ as operators, relations (\ref{eq:vz}) take the form
\begin{eqnarray*}
  {\cal V}_1 = \left(1-\beta {\partial\over\partial x_1}\right)^{-1}
        {\partial\over\partial x_1}  \\
  {\cal V}_2 = \left(1-\beta {\partial\over\partial x_2}\right)^{-1}
        {\partial\over\partial x_2}
\end{eqnarray*}

To act on polynomials, expand
$(1-\beta\; {\partial / \partial x_i})^{-1}$ in geometric series
$$
(1-\beta\; {\partial / \partial x_i})^{-1} =
\sum_{n\ge0}\beta^n\left({\partial\over\partial
    x_i}\right)^n
$$
So we have both a ${\cal V}_1$-Appell system 
and a ${\cal V}_2$-Appell system as in Section \ref{sec:apsys}.
The Appell space decompositions are, for ${\cal V}_1$ and ${\cal V}_2$,  
\begin{eqnarray*}
{\cal Z}_n^{(1)} = |\hbox{poly}_n(K)
\;\hbox{poly}\,(G)\;\Omega\:\rangle
\phantom{\biggm|} \\
{\cal Z}_n^{(2)} = |\;\hbox{poly}_n(G)\;\hbox{poly}\,(K)\;\Omega\:\rangle 
\end{eqnarray*}

respectively, where poly($\cdot$), resp. poly${}_n(\cdot)$, denote
arbitrary polynomials in the indicated variable, resp. of degree a
most $n$ in the variable. Now symmetries are generated by functions of 
${\partial / \partial x_1}$ and ${\partial / \partial x_2}$. We will
see explicit examples in Section \ref{sec:apsyss}.

\subsection{Probability distributions}

Now we shall consider some probabilistic observations. 
We introduce an inner product such that $K^*=\beta^2P_t$ and
$G^*=\beta P_x$. The $X_i$, which are formally symmetric,
extend to self-adjoint operators on appropriate domains.\\

Expectation values are taken in the state $\Omega$, i.e., for any
operator $Q$,
$$
   \langle Q\rangle_\Omega = \langle \Omega,Q\Omega\rangle
$$ 
where the normalization $\langle\Omega,\Omega\rangle=1$ is understood.  \\

From $P_t\Omega=P_x\Omega=0$ follows that
$\langle P_t\rangle_\Omega=\langle P_x\rangle_\Omega=0$ and moving $K$ and $G$
across in the inner product, that
$\langle K\rangle_\Omega=\langle G\rangle_\Omega=0$ as well.
Going back to equation (\ref{eq:gaction}), take the inner product on
the left with $\Omega$. The
 exponential factors in $K$ and $G$ average to 1, yielding
$$
    \langle e^{z_1X_1}e^{z_2X_2}\rangle_\Omega 
           = (1-\beta z_1)^{-c}
              \exp\Big({m\over 2}\,{\beta z_2^2\over 1-\beta z_1}\Big)
$$
This result has an interesting probabilistic interpretation 
for positive values of $\beta$ and $c$.
Observe that the marginal distribution of $X_1$  (i.e., for $z_2=0$)
is gamma distribution, while  the marginal distribution of $X_2$ 
(now $z_1=0$) is Gaussian. Note, however, that these are not
independent random variables.
\\

To recover the joint distribution of $X_1,X_2$, let us first recall some
probability integrals (Fourier transforms):  

\begin{eqnarray*}
\int_{-\infty}^\infty e^{i\xi y}e^{-\lambda y} 
y^{t-1}\lambda^t \,\theta(y)\,dy/\Gamma(t) &=&
(1-i\xi/\lambda)^{-t},\,{\rm for\ } t>0\\
\int_{-\infty}^\infty e^{-i\eta u}e^{-u^2/(2v)}\,du &=& \sqrt{2\pi v}\,
e^{-\eta^2v/2},\,{\rm for\ } v>0
\end{eqnarray*} 

where $\theta(x)$ denotes the usual Heaviside function, 
$\theta(x)=1$ if $x\ge0$, zero otherwise.
Replacing $z_1,z_2$ by $iz_1,iz_2$ respectively and 
taking inverse Fourier transforms, we have

\begin{proposition}
The joint density $p(x_1,x_2)$ of the random variables $X_1,X_2$ is given by
$$
 p(x_1,x_2)=
      e^{-x_1/\beta}
        \left(x_1-\frac{x_2^2}{2m}\right)^{\dot c-1}
        \beta^{-\dot c} \,
        \theta\left(x_1-\frac{x_2^2}{2m}\right)\,
        \frac{dx_1\,dx_2}{\Gamma(\dot c)\sqrt{2\pi m\beta}} 
$$
for $c,\beta>0$, where $\dot c=c-1/2$.
\end{proposition}

In the first factor, writing $x_1=x_1-x_2^2/(2m)+x_2^2/(2m)$ shows
where the Gaussian factor comes in. The result says that the marginal
distribution of $X_2$ is Gaussian with mean 0 and variance
$2m\beta$. Conditional on $X_2$, $X_1$ is gamma with parameters
$1/\beta$ and $c-1/2$ taking values in the interval
$(x_2^2/(2m),\infty)$. In the special case $c=1/2$, i.e., $\dot c=0$,
the gamma density collapses to a delta function:
$\delta(x_1-x_2^2/(2m))$.
\\

\section{Leibniz function and orthogonal basis}

Once the Leibniz formula for our Lie algebra ${\cal S}_1$ is known 
(Proposition \ref{prop:Leibnizformula}), we can proceed to define coherent 
states, find the Leibniz function --- inner product of coherent states --- 
and show that we have a Hilbert space with self-adjoint 
commuting operators $X_1=P_t+D+K$  and $X_2=G+P_x$ 
(here the $\beta$ in equations (\ref{eq:xs}) is set equal to 1). 
We recover the raising and lowering operators as elements of the
Lie algebra acting on the Hilbert space with basis consisting of the
canonical Appell system. \\

The two-parameter family of coherent states is defined as
$$
      \psi_V=\psi_{V_1,V_2}=e^{V_1K}e^{V_2G}\Omega
$$
Using Proposition \ref{prop:Leibnizformula}, we see

\begin{proposition}
\label{prop:upsilon}
With $K^*=P_t$ and $G^*=P_x$, the Leibniz function is
$$
     \Upsilon_{BV}
           = (1-B_1V_1)^{-c}
        \exp\Big({m\over 2}\,
               {B_1V_2^2+2B_2V_2+B_2^2V_1\over 1-B_1V_1}\Big)
$$
\end{proposition}

\begin{proof}
Use Proposition \ref{prop:Leibnizformula} in the relation
$$\Upsilon_{BV}= \langle \psi_B,\psi_V\rangle = \langle \Omega, 
e^{B_2P_x} e^{B_1 P_t}\,e^{V_1K}e^{V_2G}\Omega\rangle $$
\end{proof}

Note that the Leibniz function is symmetric in $B$ and $V$, which is equivalent 
to the inner product being symmetric, and thus the Hilbert space
being well-defined.
\\

It is remarkable that the Lie algebra can be reconstructed from 
the Leibniz function $\Upsilon_{BV}$.
The idea is that differentiation $\Upsilon_{BV}$ with respect to $V_1$ 
brings down $K$ acting on $\psi_V$, 
while differentiation with respect to $B_1$ brings down a
$K$ acting on $\psi_B$ which moves across the inner product as $P_t$
acting on $\psi_V$. Similarly  for $G$ and $P_x$.  We thus introduce
{\em canonical bosons}, creation operators ${\cal R}_i$, and annihilation
(velocity) operators ${\cal V}_i$,
satisfying $[{\cal V}_i,{\cal R}_j]=\delta_{ij}$.
We thus identify $K={\cal R}_1$, $G={\cal R}_2$. 
Note, however, that ${\cal V}_1$ is not the adjoint of ${\cal R}_1$, nor
${\cal V}_2$ that of ${\cal R}_2$. 
In fact, our goal is to determine the boson 
realization of $P_t$ and $P_x$, the respective adjoints.
\\

Here is a method to find the boson realization.
First, one determines the partial differential equations for
$\Upsilon=\Upsilon_{BV}$:
\begin{eqnarray*}
{\partial\Upsilon\over \partial B_1}&=&
V_1^2\,{\partial\Upsilon\over\partial V_1}+
V_1V_2{\partial\Upsilon\over \partial V_2}+ cV_1\Upsilon+
{\textstyle{m\over 2}}\,V_2^2\Upsilon \\ {\partial\Upsilon\over
\partial B_2}&=& V_1\,{\partial\Upsilon\over\partial V_2}+mV_2\Upsilon
\end{eqnarray*}

Then, one interprets each multiplication by $V_i$ as the operator
${\cal V}_i$ and each differentiation by $V_i$ as the operator 
${\cal R}_i$. This gives the following action of the operators $P_x$ 
and $P_t$ on polynomial functions of $K$ and $G$:
$$
      P_x=m{\cal V}_2+{\cal R}_2{\cal V}_1\,,\qquad 
      P_t=c{\cal V}_1+{\cal R}_1{\cal V}_1^2
           + {\textstyle{m\over 2}}\,{\cal V}_2^2
           + {\cal R}_2{\cal V}_1{\cal V}_2
$$

This means that $P_x$ acts on 
$\displaystyle |jk\rangle = {\cal R}_1^j{\cal R}_2^k|00\rangle$ 
as follows
$$ 
         P_x|jk\rangle = m\,k|j,k-1\rangle+j|j-1,k+1\rangle
$$
and $P_t$ does similarly. The element $D$ is recovered via 
$$
D   = [P_t,K]
    = [c{\cal V}_1+{\cal R}_1{\cal V}_1^2+{\textstyle{m\over 2}}\,
        {\cal V}_2^2+{\cal R}_2{\cal V}_1{\cal V}_2,{\cal R}_1]
    = c+2{\cal R}_1{\cal V}_1+{\cal R}_2{\cal V}_2
$$

Summarizing, we have

\begin{theorem}
The representation of the Schr\"odinger algebra on
the Fock space $\cal F$ with basis $|jk\rangle=K^jG^k\Omega$ is
given by
\begin{eqnarray*}  
   K  &=& {\cal R}_1 \cr
   G  &=& {\cal R}_2 \cr
   P_x&=& m{\cal V}_2+{\cal R}_2{\cal V}_1    \cr 
   P_t&=& c{\cal V}_1+{\cal R}_1{\cal V}_1^2
           + {\textstyle{m\over 2}}\,{\cal V}_2^2
           + {\cal R}_2{\cal V}_1{\cal V}_2 \cr
   D &=&  c+2{\cal R}_1{\cal V}_1+{\cal R}_2{\cal V}_2 \cr
   M &=&  m
\end{eqnarray*}
\end{theorem}

\begin{corollary}
\label{co:schr}
In the above representation, the Schr\"odinger operator
${\bf S}=P_t-P_x^2/(2m)$ is represented by
$$
   {\bf S} = \dot c{\cal V}_1 + {\cal R}_o{\cal V}_1^2
$$
where we define ${\cal R}_0={\cal R}_1 - {\cal R}_2^2/(2m)$,
\emph{cf.} Theorem \ref{thm:stdform}.
\end{corollary}

A very important feature of the Leibniz function $\Upsilon_{BV}$ 
is that it is the generating function for the inner products 
of the elements of the basis. 
Indeed, expanding the exponentials defining the coherent states yields
$$ 
\Upsilon_{BV} = \sum_{j,k,j',k'} 
                \langle jk|j'k'\rangle 
                \frac{B_1^jB_2^kV_1^{j'}V_2^{k'}}{j!k!j'!k'!}
$$
For an orthogonal basis, a necessary and sufficient condition is 
that this must be a function only of the pair
products $B_1V_1$ and $B_2V_2$. We proceed to find an orthogonal
basis.

\begin{lemma}
\label{lem:OB}
The Leibniz function can be expressed as
$$
\Upsilon_{BV} 
      = (1-B_1V_1)^{-\dot c}\,
        \exp\left(\frac{B_1}{2m}\,\frac{\partial^2}{\partial B_2^2} 
                 +\frac{V_1}{2m}\,\frac{\partial^2}{\partial V_2^2}\right)\,
        e^{mB_2V_2}
$$
with $\dot c=c-1/2$.
\end{lemma}

\begin{proof}
In the formulation of Corollary \ref{cor:leib} first set
$B_2=0$. Then use the special realization as in equation
(\ref{eq:special}) with $x=B_2$. As in Corollary \ref{cor:leib}

\begin{eqnarray*}
&&\exp\left(\frac{B_1}{2m}\frac{d^2}{dB_2^2}\right)
       \exp\left(V_1\frac{mB_2^2}{2}+V_2\,mB_2\right)\\
&&=\exp\left(\frac{V_1}{1-B_1V_1}\frac{mB_2^2}{2}+\frac{V_2}{1-B_1V_1}\,
       mB_2\right)\\
&&\qquad\times\,(1-B_1V_1)^{-1/2}
       \exp\left(\frac{m}{2}\,{\frac {B_1\,V_2^2}{1-{ B_1}\,{ V_1}}}\right)
\end{eqnarray*}

which combines to yield $\Upsilon_{BV}$ up to the factor
$(1-B_1V_1)^{-\dot c}$. Now observe that
$$ 
e^{mB_2V_2+(m/2)V_1B_2^2}
      =\exp\left(\frac{V_1}{2m}\,
       \frac{\partial^2}{\partial V_2^2}\right)\,e^{mB_2V_2}
$$
where on $\exp(mB_2V_2)$, $\partial/\partial V_2$ acts simply as
multiplication by $mB_2$. Combining with the above observations yields
the result.
\end{proof}

Now for the main result 
(expressed in terms of $R_0=K-G^2/(2m)$, see Corollary \ref{co:schr})

\begin{theorem} 
The set $|ab\rangle=R_0^aG^b\Omega$, $a,b\ge0$, forms an
orthogonal basis with squared norms equal
$$
    \langle ab|ab\rangle =  (\dot c)_a\,a!b!m^b 
$$
where $(\dot c)_a=\dot c(\dot c+1)\cdots(\dot c+a-1)$.
\end{theorem}

\begin{proof}
From Lemma \ref{lem:OB}, 
\begin{eqnarray*}
  &&(1-B_1V_1)^{-\dot c}\, e^{mB_2V_2}\\
  &&\quad=\exp\left(-\frac{B_1}{2m}\,
    \frac{\partial^2}{\partial B_2^2}
          -\frac{V_1}{2m}\,\frac{\partial^2}{\partial V_2^2}\right)\,
    \langle e^{B_1K+B_2G}\Omega,e^{V_1K+V_2G}\Omega \rangle \\
&&\quad=
    \langle e^{B_1(K-G^2/(2m))+B_2G)}
       \Omega,e^{V_1(K-G^2/(2m))+V_2G}\Omega\rangle\\
&&\quad = 
     \langle e^{B_1R_0+B_2G}\Omega,e^{V_1R_0+V_2G}\Omega \rangle
\end{eqnarray*}

Now we have the generating function for the inner products
$\langle ab|a'b' \rangle$ depending only on the pair products $B_1V_1$,
$B_2V_2$. Hence orthogonality. Expanding the left-hand side of the
equation yields the squared norms.
\end{proof}

Similarly, we have for the canonical Appell system,

\begin{proposition}
  Let $X_0=X_1-X_2^2/(2m)$, with the identification \hfill\break
  $X_0\Omega=x_0\cdot 1$.
Then 
$$
      e^{v_0R_0}e^{v_2G}\Omega=\exp(x_0\,{v_0\over 1+\beta v_0}) 
       (1+\beta v_0)^{-\dot c}\exp(x_2v_2-\beta mv_2^2/2)
$$
\end{proposition}

\begin{proof}
  First substitute $v_0$ for $v_1$ in Proposition
  \ref{prop:genfun}. And observe that
$$
    e^{v_0R_0}e^{v_2G}\Omega 
       = \exp\left(-\frac{v_0}{2m}\,
         \frac{\partial^2}{\partial v_2^2}\right)e^{v_0K}e^{v_2G}\Omega
$$
Now use the special realization, equation (\ref{eq:special}), taking
$x=v_2$ in Corollary \ref{cor:leib}, with
$$
       B_1=-v_0,\quad 
       B_2=0,\quad 
       V_1=-\frac{\beta}{1+\beta v_0},\quad
       V_2=\frac{x_2/m}{1+\beta v_0}
$$
After substituting accordingly and simplifying, one finds the stated result.
\end{proof}

Note that now the system decouples into Laguerre polynomials in the
variable $x_0$ and Hermite polynomials in the variable $x_2$.
\\

\section{Evolution equations and Appell systems}
\label{sec:apsyss}

We shall now consider evolutions on the 2-dimensional space $M$ spanned by
$G$ and $R_0$ in terms of scalar functions $u$ on $M$
parametrized by time $\tau$.  
Note that polynomial functions on $M$ can be interpreted as vectors of the Fock space 
${\cal F}=\hbox{span}\,\{\,|jk\rangle = K^jG^k\Omega\,\}$
of the representation of the Schr\"odinger algebra.
The evolution equation is
\begin{equation}\label{eq:evol}
u_{\tau} = H u
\end{equation}
where the subscript $\tau$ denotes differentiation over the time parameter.
For generators of evolutions, we shall consider operator functions 
$H=H(P_t,P_x)$.  In order to have an operator calculus, the function $H$ 
is chosen so  that $H(\partial/\partial x_1,\partial/\partial x_2)$
generates a L\'evy process on ${\bf R}^2$, with its associated convolution 
semigroup of probability measures providing the fundamental solution
to the evolution equation (\ref{eq:evol}).
Taking such a process $w(\tau)=(w_1(\tau),w_2(\tau))$
with generator  $H$,
consider the action of the group element $\exp(w_1(\tau)P_x+w_2(\tau)P_t)$
as an operator on functions of $K$ and $G$. Then the (partial) group
law yields the solution to the evolution equation $u_{\tau}=Hu$ with
elements of the basis $|jk\rangle = K^jG^k\Omega$ as initial
functions. These are Appell systems on the Schr\"odinger algebra. In this
case,
since the elements $P_x$ and $P_t$ commute, the process considered on
the group is essentially the same as that in Euclidean space, and 
we have an operator calculus for the operators $P_t$ and $P_x$ via
the generator of the process $H$.\\

As we have seen in the previous section, it is convenient to use $R_0$
instead of $K$, so we will take as initial functions
$|ab\rangle=R_0^aG^b\Omega$.
A special case of Proposition \ref{prop:Leibnizformula}, with
$B_2=V_2=0$ gives the Leibniz formula for sl(2), namely
$$e^{w_1L_0}e^{v_0R_0}=\exp\left(\frac{v_0}{1-v_0w_1}\,R_0\right)
(1-v_0w_1)^{-\rho_0}\exp\left(\frac{w_1}{1-v_0w_1}\,L_0\right)$$
Similarly, taking in the same formula $B_1=V_1=0$, yields the Leibniz
formula for the HW-algebra, which we want in the form
$$\exp(w_2P_x)\exp(v_2G)=\exp(m\,w_2v_2)\exp(v_2G)\exp(w_2P_x)$$
Thus,
\begin{proposition}
  The abelian subgroup generated by $P_t$ and $P_x$ on the basis
  $|ab\rangle$ has generating function
\begin{eqnarray*}
&&e^{w_1P_t+w_2P_x}e^{v_0R_0+v_2G}\Omega=\\
&&\qquad\exp\left(\frac{v_0}{1-v_0w_1}\,R_0\right)
(1-v_0w_1)^{-\dot c}
\exp\left(v_2(G+mw_2)+mw_1v_2^2/2\right)
\end{eqnarray*}
\end{proposition}

\begin{proof}
First, use the fact that $P_x$ commutes with $R_0$ together with the
Leibniz formula for the HW-algebra to yield
$$ 
     e^{w_1P_t}e^{v_0R_0}e^{v_2G}e^{m\,w_2v_2}\Omega 
$$
Now write, using the standard form, Theorem \ref{thm:stdform},
$L=L_0+L_1$, with $L_1=P_x^2/(2m)$. Since $L_1$ commutes with $R_0$,
we have
$$ 
    e^{w_1L_0}e^{v_0R_0}e^{w_1L_1}e^{v_2G}e^{m\,w_2v_2}\Omega 
$$
We can reconstitute $L_1$ as $P_t$ since $L_0$ commutes with $G$, and
use Proposition \ref{prop:Leibnizformula}, or, equivalently, think of
$P_x$ as acting formally like $\displaystyle m\,\frac{d}{dG}$. In any case, we
arrive at
$$ 
     e^{w_1L_0}e^{v_0R_0}e^{w_1L_1}e^{v_2G}e^{m\,w_2v_2+m\,w_1v_2^2/2}\Omega
$$
Now use the Leibniz formula for sl(2) mentioned above, and since the
$\rho_0$ and $L_0$ commute with $G$, and $\rho_0\Omega=\dot c\Omega$, 
the stated result follows.
\end{proof}

Finally, the solution to the evolution equation $u_{\tau}=Hu$ is found by
averaging over $w_1$ and $w_2$, i.e., the angle brackets on the
right-hand side denoting expected value,
$$
    e^{\tau\,H(P_t,P_x)}\,|ab\rangle 
        = \langle \exp\left(\frac{v_0}{1-v_0w_1}\,R_0\right)
         (1-v_0w_1)^{-\dot c}
         \exp\left(v_2(G+m\,w_2)+m\,w_1v_2^2/2\right) \rangle
$$
Expanding in $v_0,v_2$ and evaluating moments of the process yields
the corresponding Appell systems.\\

For $w_1,w_2$ independent Gaussians with mean zero and variance
$\tau$, we have $H(P_t,P_x)=\half\,(P_t^2+P_x^2)$, with $u_{\tau}=Hu$ 
a natural extension of the classical heat equation.
\\


\end{document}